\documentclass[12pt]{article}
\usepackage{amssymb}
\usepackage{epsfig}

\newcommand{\beqn}{\begin{eqnarray}}
\newcommand{\eeqn}{\end{eqnarray}}
\newcommand{\beq}{\begin{equation}}
\newcommand{\eeq}{\end{equation}}

\newcommand{\tr}{\mbox{Tr}}

\begin{document}
~ \vspace{-1cm}
\begin{flushright}
{ ITEP-LAT/2004-04}

\vspace{0.2cm}

{ MPI-2004-9}

\end{flushright}

\begin{center}
{\baselineskip=24pt {\Large \bf A novel probe of the vacuum of the
lattice gluodynamics}\\

\vspace{1cm}

{\large
    M.\,I.\,Polikarpov$^{\dag}$,
    S.\,N.\,Syritsyn$^{\dag}$ and
    V.\,I.\,Zakharov$^{*}$} } \vspace{.5cm} {\baselineskip=16pt { \it

$^{\dag}$ Institute of Theoretical and  Experimental Physics,
B.~Cheremushkinskaya~25, Moscow, 117259, Russia\\
$^{*}$ Max-Planck Institut f\"ur Physik, F\"ohringer Ring 6,
80805, M\"unchen, Germany} }
\end{center}

\vspace{5mm}

\date{}
\abstract{We introduce a notion of minimal number of negative links on the
lattice for a given original configuration of $SU(2)$ fields. Negative links
correspond to a large potential, not necessarily large action. The idea is
that the minimal number of negative links is a gauge invariant notion. To
check this hypothesis we measure correlator of two negative links, averaged
over all the directions, as function of the distance between the links. The
inverse correlation length coincides within the error bars with the lightest
glueball mass. }

\vspace{0.5cm}

{\bf 1} A traditional way to study spectrum of excitations is to
measure correlators of various sources. For example, the
correlator\footnote{For simplicity of notations, we do not
account for the anomalous dimension.}:
\beq \label{independent}
D_{gl}(r)~\equiv ~\langle 0|(G^a_{\mu\nu}(r))^2,
~(G^a_{\alpha\beta}(0))^2|0\rangle~~,
\eeq
where $G_{\mu\nu}^a$ is
the gluonic field strength tensor, is sensitive to the glueball
mass $m_{gl}$ at large (Euclidean) distances $r$:
\beq\label{mass}
\lim_{r\to\infty}D_{gl}(r)~=const~+(const)^{'}\exp(-m_{gl}r)~~.
\eeq
On the other hand, one can also study gauge dependent correlators, such as
\beq\label{dependent}
D_{\mu\nu}^{a,b}(r)~=~\langle 0|A_{\mu}^a(r),A_{\nu}^b(0)|0\rangle~~,
\eeq
where $A_{\mu}^a$ is the gauge field. Such correlators are not unitary,
generally speaking, and are not controlled by the glueball mass. In
particular, the gluon propagator (\ref{dependent}) could even grow at large
$r$. More realistically, i.e., as indicated by the lattice measurements, the
gluon propagator falls off at large $r$ but exhibits some spurious mass
scales, for review see, e.g., \cite{alkofer}. For us, it is important that
these spurious mass scales are, as a rule, lower than the glueball mass.

{\bf 2} In this note we introduce a new type of correlators
which, as we hypothesize, might be unitary although they are not
explicitly gauge invariant (like (\ref{independent})) and check
our hypothesis through lattice simulations of $SU(2)$
gluodynamics.

To explain the basic idea behind our measurements it is useful
first to remind to the reader how one can introduce a gauge
invariant condensate of dimension two in gauge theories
\cite{stodolsky}. One starts with the vacuum expectation value
$<(A_{\mu}^a)^2>$ which is obviously gauge dependent. One can,
however, minimize this vacuum expectation value on the gauge
orbits and the results $<(A_{\mu}^a)^2>_{min}$ is gauge
invariant by construction. To ensure that the minimum exists,
the Euclidean signature is used\footnote{A modification suitable
for the Minkowski space is to work in the Hamiltonian formalism
\cite{baal}.}.

We generalize this idea to the case of a $Z(2)$ projection of
the original $SU(2)$ fields. In this projection the link
variables take the values $\pm 1$ (for review see, e.g.,
\cite{greensite}). Upon the $Z(2)$ projection there is still
remaining $Z(2)$ gauge invariance. We fix this gauge freedom by
minimizing the number of negative links over the whole lattice.
We speculate, furthermore, that the density of these negative
links might well be gauge invariant, in analogy with
$<(A_{\mu}^a)^2>_{min}$.

Moreover, we expect that the correlator of the negative links is
unitary, like (\ref{independent}). This speculation goes
apparently beyond the ideas discussed so far. Indeed, the
negative links in the continuum limit correspond to singular
potentials, $A^a\sim 1/a$, where $A$ is the lattice spacing.
Naively, one could argue that such fields are artifacts of the
lattice and, as a manifestation of this, their correlator dies
off on distances of order $a$. Why we call these arguments
naive, is because the so called P-vortices are defined as
unification of all the negative plaquettes in the $Z(2)$
projection and exhibit sensitivity to the scale of
$\Lambda_{QCD}$, for review see \cite{greensite}. It is in
analogy with these observations that we expect the ``propagator
of negative links'' to scale in the physical units as well. This
is of course a daring possibility which cannot be proven a
priory but only supported or rejected by measurements.

{\bf 4} In more detail, we perform measurements both in the Direct- and
Indirect-Maximal Center Projections (DMCP and IMCP). The details of
calculations are given in the Appendix. As a result of $SU(2) \to Z(2)$
projection, the original $SU(2)$ field configurations get projected into the
closest configuration of $Z(2)$ gauge fields. The remaining $Z(2)$ gauge
freedom is then fixed by maximizing the functional
\beq
F(Z) = \sum_{x,\mu} Z_{x,\mu}
\eeq
with respect to $Z(2)$ gauge transformations ($Z_{x,\mu} \to z_x\, Z_{x,\mu}
\, z_{x+\hat{\mu}}$, $z_x=\pm 1$).

After gauge fixing only the positions of the negative links are relevant
and it is reasonable to change the variables:
\beq
\hat{Z}_{x,\mu}~=~\{1,~if~ Z_{x,\mu}=-1;~~0~if~Z_{x,\mu}=1\}~~
\eeq
Moreover, to imitate a scalar correlator on the discrete variables,
we consider here the isotropic correlator defined as
\beq\label{correlator}
G_{\mu\nu}(r)~\equiv~ \frac{1}{N_{_r}}\sum_{r < |x| <r+\frac a2}
<\hat{Z}_{0\mu}\hat{Z}_{x,\nu}>\, ,
\eeq
where the summation is over all links $Z_{x,\mu}$ for $x$ lying in the
spherical layer $r < |x| <r+\frac a2$; $N_r$ is the total number of links in
this layer.

\begin{figure}[h]
\begin{center}
\includegraphics[scale=0.5, angle=270]{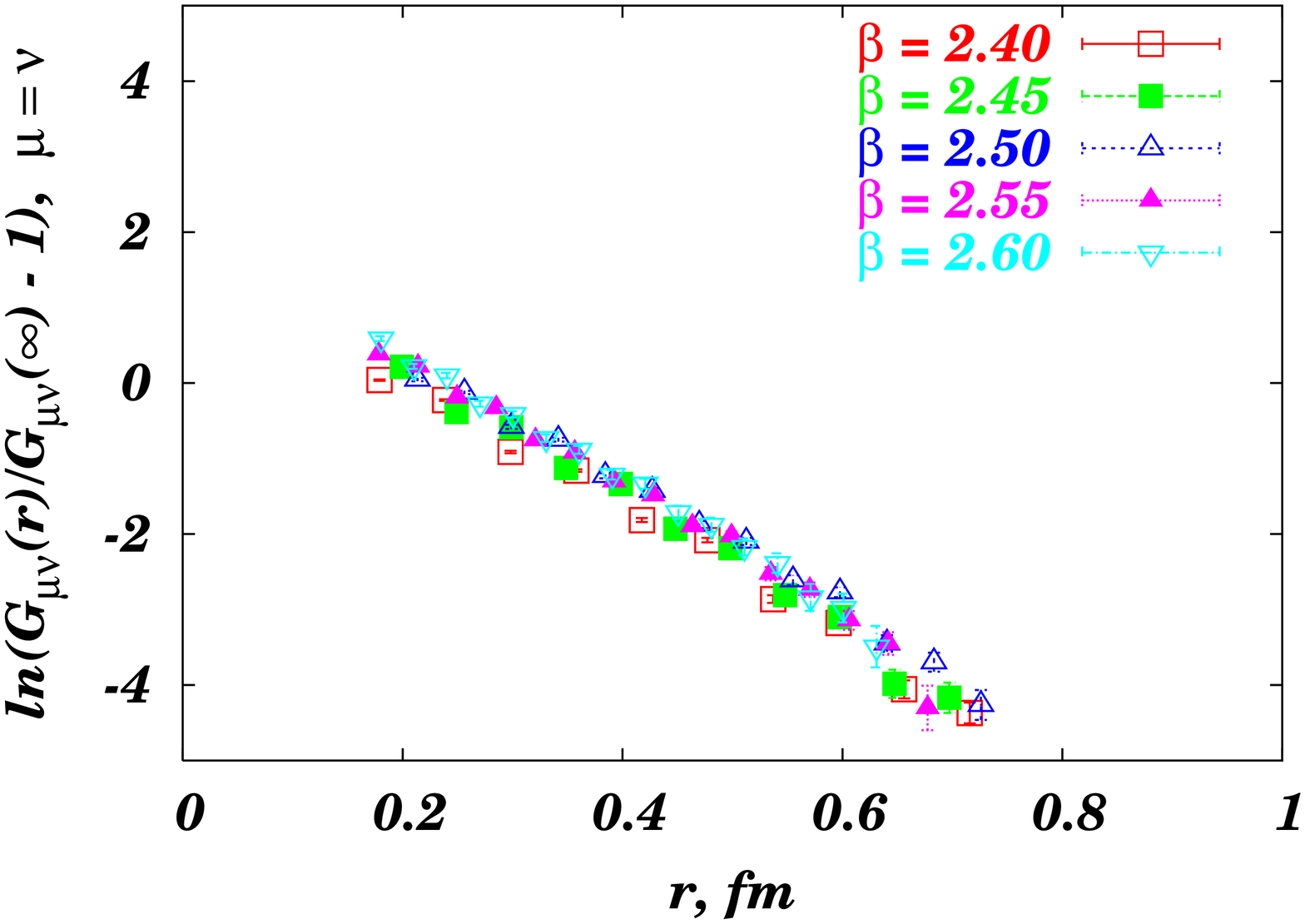}\\
(a)\\
\includegraphics[scale=0.5, angle=270]{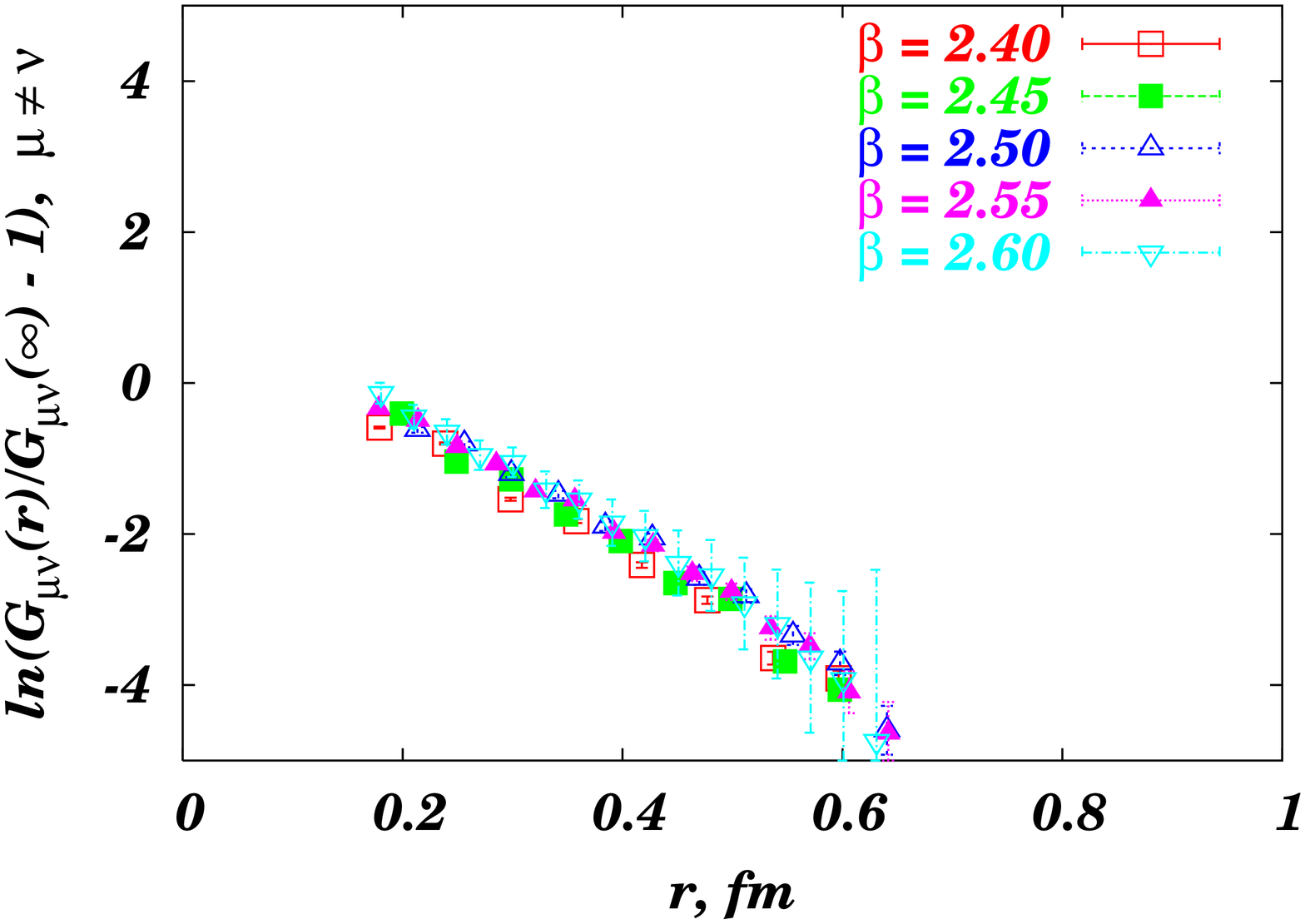}\\
(b)
\end{center}
\caption{$\ln({G_{\mu\nu}(r)\over G(\infty)}-1)$ vs lattice spacing for
$\mu=\nu$ (a) and for $\mu\neq\nu$ (b), results obtained for IMCP.
\label{fig:lnG}}
\end{figure}

\begin{figure}[h]
\begin{center}
\includegraphics[angle=270,scale=0.5]{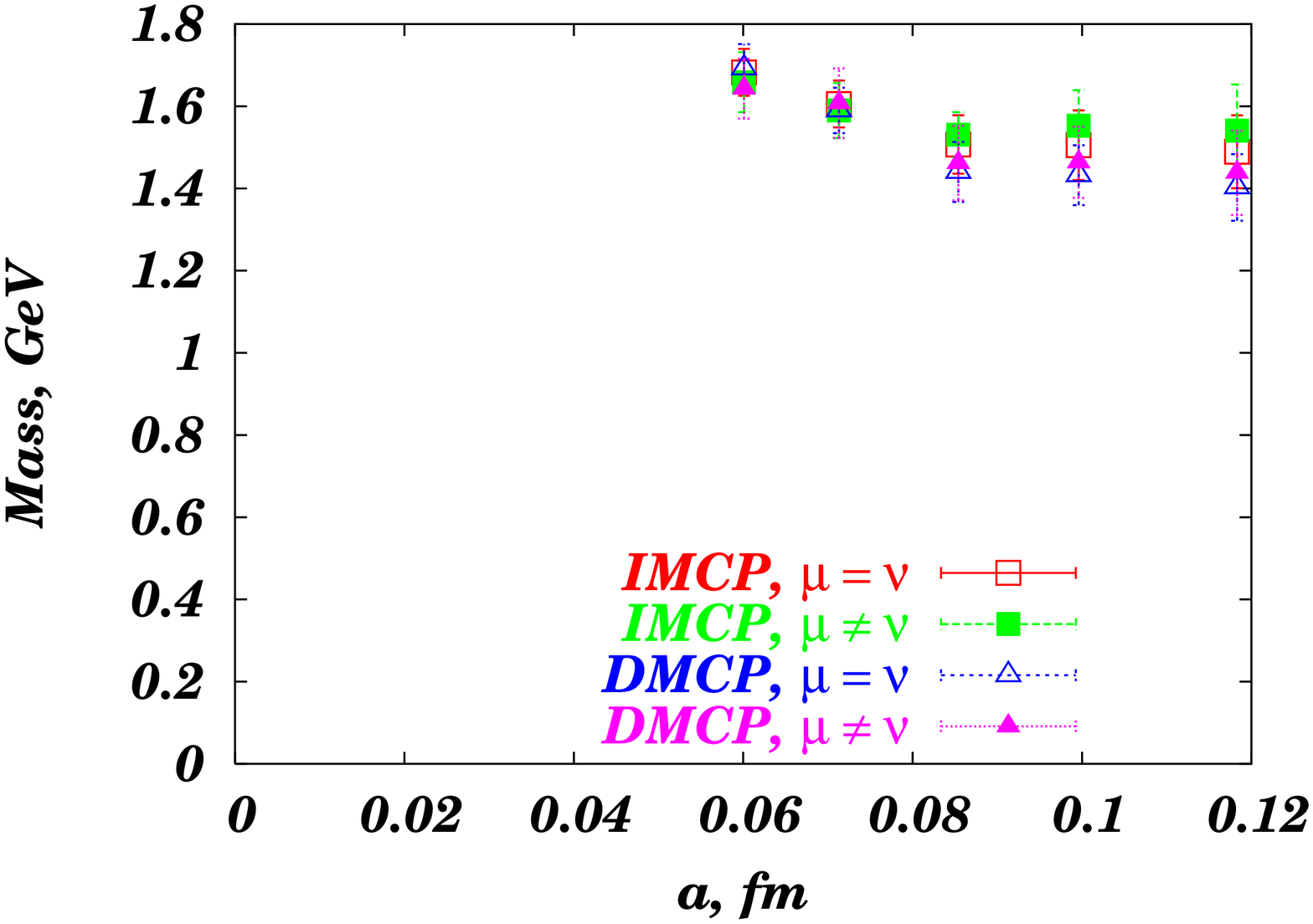}
\end{center}
\caption{Mass parameters for IMCP and DMCP for $\mu=\nu$ (a) and for
$\mu\neq\nu$. \label{fig:mass}}
\end{figure}

The correlator (\ref{correlator}) tends to a non-vanishing constant
$G(\infty) = <\hat{Z}_{x,\nu}>^2$ as $r\to \infty$ ($<\hat{Z}_{x,\nu}>$ is
the average density of the negative links) and we fit the data by the
expression (see (\ref{mass})):
\beq
G(r)~=~G(\infty)+C\exp\{-mr\} \, .
\eeq
We thus get the mass parameter $m$ for various values of the lattice spacing
$a$. Logarithmic plots for the isotropic correlator $G_(r)$ are presented in
Fig.~\ref{fig:lnG} for IMCP, while the corresponding values of the mass $m$
are depicted in Fig.~\ref{fig:mass} for IMCP and DMCP.

{\bf 5} As is seen from Figs.~\ref{fig:lnG},\ref{fig:mass} the correlator
$G_{\mu\nu}(r)$ scales in the physical units (within the error bars) and the
mass parameter is close to the scalar $0^{++}$ glueball mass ($m(0^{++})=
1.65\pm0.05 \, Gev$~\cite{Teper:1998kw}). Thus, our measurements support the
idea that the correlator (\ref{correlator}) is in fact gauge invariant and
unitary. Of course the results are not analytical but pure numerical and, in
principle, the picture can change at smaller values of the lattice spacing.
The non-triviality of this observation is that it is a correlator of
potentials, which are not explicitly gauge invariant. Moreover, in the
continuum limit the negative links correspond to singular potentials.

\section*{Acknowledgements} We would like to thank V.G.~Bornyakov and
G.~Greensite for very useful discussions. M.I.P. and S.N.S. are partially
supported by grants RFBR 02-02-17308, RFBR 01-02-17456, DFG-RFBR 436 RUS
113/739/0, INTAS-00-00111, and CRDF award RPI-2364-MO-02. V.I.Z. is partially
supported by grant INTAS-00-00111.

\section*{Appendix}

We perform our calculations both in the Direct~\cite{greens2} -- and the
Indirect~\cite{debbio} -- Maximal Center Projections (DMCP and IMCP). The
DMCP in SU(2) lattice gauge theory is defined by the maximization of the
functional
\begin{equation}
F_1(U) = \sum_{x,\mu} \left( \tr U_{x,\mu}\right)^2 \, , \label{maxfunc}
\end{equation}
with respect to gauge transformations, $U_{x,\mu}$ is the lattice gauge
field. Maximization of (\ref{maxfunc}) fixes the gauge up to Z(2) gauge
transformations and the corresponding Z(2) gauge field is defined as:
$Z_{x,\mu} = \mbox{sign} \tr U_{x,\mu}$. To get IMCP we first fix the
maximally Abelian gauge maximizing the functional
\begin{equation}
F_2(U) = \sum_{x,\mu} \tr \left( U_{x,\mu}\sigma_3 U_{x,\mu}^+
\sigma_3\right) \, , \label{maxfuncmaa}
\end{equation}
with respect to gauge transformations. We project gauge degrees of freedom
$U(1)\to Z(2)$ by the procedure completely analogous to the DMCP case, that
is we maximize $F_1(U)$ (\ref{maxfunc}) with respect to $U(1)$ gauge
transformations.

To fix the maximally Abelian and direct maximal center gauge we create 20
randomly gauge transformed copies of the gauge field configuration and apply
the Simulated Annealing algorithm to fix gauges. We use in calculations that
copy which correspond to the maximal value of the gauge fixing functional. To
fix the indirect maximal center gauge from configuration fixed to maximally
Abelian gauge and to fix the Z(2) degrees of freedom one gauge copy is enough
to work with our accuracy. We work at various lattice spacings to check the
existence of the continuum limit. The parameters of our gauge field
configurations are listed in Table~\ref{conf_table}. To fix the physical
scale we use the string tension in lattice units~\cite{Fingberg:1992ju},
$\sqrt\sigma = 440\, MeV$.

\begin{table}[t]\label{conf_table}
\caption{Parameters of configurations.}
\begin{center}
\begin{tabular}{|c|c|c|c|}
\hline
$\beta$ & Size &    $N_{IMCP}$ &    $N_{DMCP}$ \\
\hline
$2.35$ &    $16^4$ &    $20$ &  $20$ \\
$2.40$ &    $24^4$ &    $50$ &  $20$ \\
$2.45$ &    $24^4$ &    $20$ &  $20$ \\
$2.50$ &    $24^4$ &    $50$ &  $20$ \\
$2.55$ &    $28^4$ &    $37$ &  $17$ \\
$2.60$ &    $28^4$ &    $50$ &  $20$ \\
\hline
\end{tabular}
\end{center}
\end{table}

\end{document}